\documentclass[aps,prl,onecolumn,11pt,showpacs,superscriptaddress,amsmath,amssymb,nobibnotes,letterpaper]{revtex4-1}

\usepackage{blindtext}  
\usepackage{amssymb}
\usepackage{mathtools}
\usepackage[pdftex]{graphicx}
\usepackage[usenames, dvipsnames, svgnames, table]{xcolor}
\usepackage[colorlinks=true, citecolor=Blue,
            linkcolor=BrickRed, urlcolor=ForestGreen]{hyperref}
\usepackage{txfonts}
\usepackage{mathrsfs}
\usepackage{bm}
\usepackage{multirow}
\usepackage{color}
\usepackage{comment}
\usepackage{soul}

\usepackage{algorithm}  
\usepackage{algpseudocode}
\begin{document}

\newcommand{\Com}[1]{{\color{red}{#1}\normalcolor}} 


\title{PPO-Q: Proximal Policy Optimization with Parametrized Quantum Policies or Values}

\author{Yu-Xin Jin}
\email{These authors contribute equally}
\affiliation{Beijing Academy of Quantum Information Sciences, Beijing 100193, China}

\author{Zi-Wei Wang}
\email{These authors contribute equally}
\affiliation{Beijing Academy of Quantum Information Sciences, Beijing 100193, China}

\author{Hong-Ze Xu}
\affiliation{Beijing Academy of Quantum Information Sciences, Beijing 100193, China}

\author{Wei-Feng Zhuang}
\affiliation{Beijing Academy of Quantum Information Sciences, Beijing 100193, China}

\author{Meng-Jun Hu}
\email{humj@baqis.ac.cn}
\affiliation{Beijing Academy of Quantum Information Sciences, Beijing 100193, China}

\author{Dong E. Liu}
\email{dongeliu@mail.tsinghua.edu.cn}
\affiliation{State Key Laboratory of Low Dimensional Quantum Physics, Department of Physics, Tsinghua University, Beijing, 100084, China}
\affiliation{Frontier Science Center for Quantum Information, Beijing 100184, China}
\affiliation{Beijing Academy of Quantum Information Sciences, Beijing 100193, China}
\affiliation{Hefei National Laboratory, University of Science and Technology of China, Hefei, 230088, China}

\begin{abstract}

Quantum machine learning (QML), which combines quantum computing with machine learning, is widely believed to hold the potential to outperform traditional machine learning in the era of noisy intermediate-scale quantum (NISQ). As one of the most important types of QML, quantum reinforcement learning (QRL) with parameterized quantum circuits as agents has received extensive attention in the past few years. Various algorithms and techniques have been introduced, demonstrating the effectiveness of QRL in solving some popular benchmark environments such as CartPole, FrozenLake, and MountainCar. However, tackling more complex environments with continuous action spaces and high-dimensional state spaces remains challenging within the existing QRL framework. Here we present PPO-Q, which, by integrating a hybrid quantum-classical networks into the actor or critic part of the proximal policy optimization (PPO) algorithm, achieves state-of-the-art performance in a range of complex environments with significantly reduced training parameters. The hybrid quantum-classical networks in the PPO-Q incorporates two additional traditional neural networks to aid the parameterized quantum circuits in managing high-dimensional state encoding and action selection. When evaluated on $8$ diverse environments, including four with continuous action space, the PPO-Q achieved comparable performance with the PPO algorithm but with significantly reduced training parameters. Especially, we accomplished the BipedalWalker environment, with a high-dimensional state and continuous action space simultaneously, which has not previously been reported in the QRL. More importantly, the PPO-Q is very friendly to the current NISQ hardware. We successfully trained two representative environments on the real superconducting quantum devices via the Quafu quantum cloud service. Our work paves the way for exploring more sophisticated control problems via QRL.

\end{abstract}

\maketitle

\section{1. Introduction} 
\label{sec1}
Quantum computing is a completely different computing paradigm, which has been proven to provide significant computational advantages over classical computing in some important fields such as prime factorization and search problem \cite{bib16}. However, the realization of universal quantum computing is still in its infancy and has a long journey ahead. At present, we are in the era of noisy intermediate-scale quantum (NISQ), which indicates that the qubit counts, qubit quality and gate operation fidelity are all constrained \cite{bib17, bib14, bib15}. The variational quantum algorithms, which utilize low-depth parameterized quantum circuits and classical optimization subroutines to adjust trainable parameters, are widely acknowledged as the promising algorithms endowed with application potential and exhibiting a high degree of suitability for the current NISQ hardwares \cite{bib18, bib19, bib22, bib23}. As the prime example of variational quantum algorithms, quantum machine learning (QML) \cite{bib20, bib21} has drawn extensive attention and research because of its potential advantages in the field of artificial intelligence \cite{bib24,bib25,bib26,bib27,bib28,bib29,bib30, bib31, bib32, bib33, bib34, bib6,bib12,bib35,bib36}.

Reinforcement learning (RL), a different paradigm within the realm of machine learning, involves an agent learning to optimize its behavior over time to maximize cumulative rewards via interaction with the environment.  RL has attained remarkable achievements in the past decade across a diverse range of applications, such as triumphing over the world champion in the game of Go \cite{bib38}, demonstrating excellent performance in the game of StarCraft \cite{bib39}, and resolving complex real-world conundrums \cite{bib40,bib41,bib42,bib43}. Among the various algorithms in the field of RL, proximal policy optimization (PPO) algorithm \cite{bib44} has received special attention because of its successful application in various domains such as robotics, gaming, and autonomous driving. PPO combines policy network (actor) and value network (critic), 
its effectiveness and versatility make it a popular choice \cite{bib13} for solving complex decision-making and control problems, significantly advancing the capabilities and applications of RL. 
Given the potential superiority of quantum computing, it is natural to investigate whether quantum reinforcement learning (QRL), which combines quantum computing and reinforcement learning, is capable of yielding unexpected advantages.

Quantum reinforcement learning, which utilizes parameterized quantum circuit (PQC) as the agent, has received lots of attention in both value-based \cite{bib12, bib35} and policy-based \cite{bib6, bib45, bib46} RL methods in the past few years. These approaches have successfully addressed some popular benchmark environments such as CartPole, FrozenLake, and MountainCar from OpenAI Gym \cite{brockman2016openai}, thereby demonstrating the effectiveness of QRL. In order to tackle more complex environments or control problems, more advanced algorithms \cite{bib8, bib9, bib10} and architecture design techniques \cite{bib5, bib7, bib11, bib47, bib49} have been introduced in the realm of QRL. However, existing QRL framework still struggle to address more complex benchmark environments with continuous action spaces and high-dimensional state spaces, not to mention accomplishing real-world tasks.

\begin{figure}[htbp]
    \centering
    \includegraphics[width=1\linewidth]{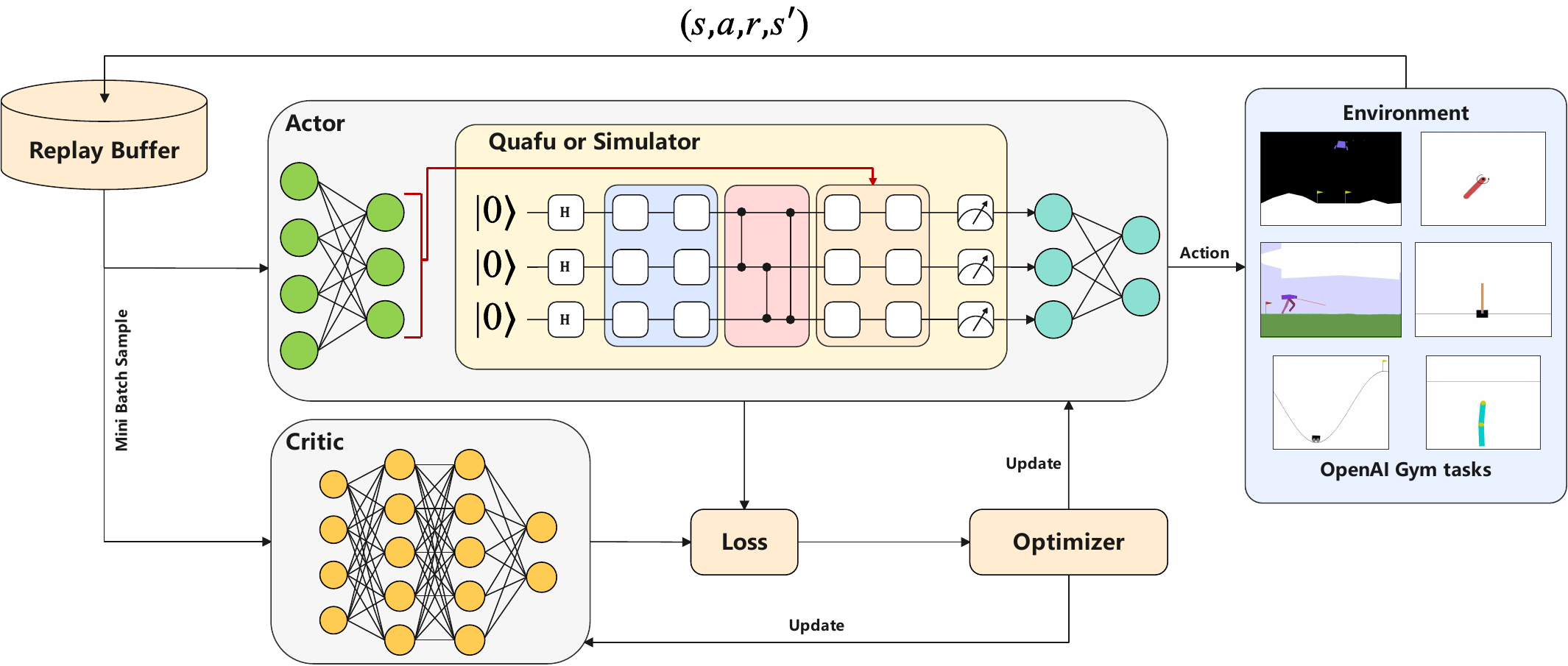}
    \caption{The framework of PPO-Q, which provides an example of assigning the actor to PQC and the critic to classical NNs. The quantum actor interacts with various environments from OpenAI Gym \cite{brockman2016openai}, filling the replay buffer with a set of data $(s,a,r,s^{\prime})$. A mini-batch is then sampled to update parameters within PPO algorithm.}
    \label{fig1}
\end{figure}

Here we present PPO-Q, which is a general hybrid quantum-classical RL framework based on the PPO algorithm, to address more complex environments with continuous action spaces and high-dimensional state spaces. The core of PPO-Q, as shown in Fig.\ref{fig1}, is a hybrid quantum-classical networks that incorporates two additional neural networks to aid the PQC in managing high-dimensional state encoding and action selection. The hybrid quantum-classical networks is incorporated into either the actor or critic component of the PPO algorithm, thereby attaining state-of-the-art performance across a variety of complex environments. When evaluated on $8$ diverse environments, including four with continuous action space, the PPO-Q achieved comparable performance with the PPO algorithm but with significantly reduced training parameters. Especially, we achieved success in the BipedalWalker environment, which simultaneously features a high-dimensional state and a continuous action space that has not been previously reported in the realm of QRL. 
More importantly, the PPO-Q exhibits high compatibility with the current NISQ hardware. Through the Quafu quantum cloud service, we have successfully trained two representative environments CartPole and LunarLander on real superconducting quantum devices, thereby demonstrating the practical applicability and effectiveness of the PPO-Q in the NISQ era. The major contributions of this work can be summarized as follows:
\begin{itemize}
    \item We present the PPO-Q, a general hybrid quantum-classical framework, which is capable of addressing  complex environments with high-dimensional state spaces and continuous action spaces that are highly challenging in the field of QRL.
    The integration of quantum and classical architectures within PPO-Q presents a novel and innovative methodology for probing the potential quantum advantage in the context of complex and challenging control problems. 
    \item We introduce several innovative techniques, including the reduction of input high-dimensional states aided by Neural Network (NNs), the use of NNs for post-processing quantum circuit readout expectations, and code-level joint optimizations. Such methods not only significantly augment performance but also make contributions to enhanced generality and robustness across diverse environments.
    \item We conduct a comprehensive evaluation of PPO-Q across multiple diverse environments, leveraging both quantum simulator and real superconducting quantum devices. The obtained results convincingly validate the efficacy of proposed method and its suitability for application in current NISQ devices, which lay a solid foundation for further research and development in the field of QRL.
\end{itemize}

\section{2. Related Works} 
\label{sec2}
Previous studies have explored various aspects of QRL, which include the development of value-based and policy-based methods, the improvement of architectural design to solve complex environments, and demonstrations on real quantum devices. A comprehensive overview of the current research landscape in QRL is available in Meyer et al. \cite{bib50}. TABLE \ref{tab01} provides an intuitive comparison of selected representative works with our work. \par
\begin{table}[htbp]
    \caption{Related works of PQC-based quantum reinforcement learning.}
    \footnotesize
    \begin{ruledtabular}
        \begin{tabular}{cccccc}
        Literature & Environments & Learning Algorithm & Solving Tasks & \begin{tabular}[c]{@{}c@{}}Comparing\\ with Classic NNs\end{tabular} & \begin{tabular}[c]{@{}c@{}}Using\\ Real Device\end{tabular} \\ \hline
        \cite{bib6} & CartPole, MountainCar, Acrobot & Policy gradient with baseline & Yes & None & No \\
        \cite{Jin_2024} & CartPole & Policy gradient with baseline & Yes & None & Yes \\
        \cite{bib12} & CartPole, FrozeLake & Deep Q-Learning & Yes & Comparable & No \\
        \cite{bib8} & Pendulum & Soft Actor-Critic & Yes & Comparable & No \\
        \cite{bib10} & CartPole, FrozeLake & Advantage Actor-Critic & Yes & Comparable & No \\
        \cite{bib7} & Pendulum, LunarLander(C) & PPO & Yes & Comparable & No \\
        \cite{kölle2024study} & Cart Pole, FrozeLake & PPO & Yes & Comparable & No \\
        
        This work & \begin{tabular}[c]{@{}c@{}}CartPole, MountainCar, Acrobot, \\ LunarLander, MountainCar(C), Pendulum, \\ LunarLander(C), BipedalWalker\end{tabular} & PPO & Yes & Comparable & Yes
        \end{tabular}
        \label{tab01}
    \end{ruledtabular}
\end{table}

One of the earliest works proposing the utilization of PQC to approximate the action-value function is presented in Chen et al. \cite{bib35}. Skolik et al. \cite{bib12} have addressed the FrozenLake and CartPole environments through quantum Q-learning agents, exploiting data re-uploading and output scaling techniques. Employing similar techniques, Jerbi et al. \cite{bib6} have further developed a quantum policy gradient algorithm with custom-designed observables to handle the CartPole, MountainCar, and Acrobot environments. Based on the quantum policy gradient algorithm, Jin et al, \cite{Jin_2024} have demonstrated the Cartpole training and inference on the real superconducting quantum hardwares.
Additionally, in an effort to resolve complex environments, Kölle et al. \cite{bib10} have proposed a quantum advantage actor-critic algorithm. Moreover, quantum versions of the soft actor-critic algorithm \cite{bib8, bib9}, assisted by classical neural networks, have been proposed to solve the Pendulum with a continuous action space and to control a virtual robotic arm.\par

In the past few years, the typical RL algorithm PPO has attracted significant attention within the realm of QRL, with expectation that its quantum variants hold potential to tackle complex problems more efficiently. Kwak et al. \cite{bib48} have initially substituted the actor network with a PQC in PPO algorithm; nevertheless, they failed to effectively solve the simple CartPole problem. To address trainability, a scheme of unentangled quantum agent using one-qubit circuits was proposed \cite{bib11}, which successfully trained the CartPole, Acrobot, and LunarLander environments and was verified on real quantum hardware. However, unentangled quantum agent  does not have generalization performance.
Meanwhile, careful design choices for PQCs, such as angle embedding, encoding block architecture, and post-processing, have been refined for continuous environments \cite{bib7}. Additionally, other optimization techniques like exponential learning
rate decay have also been considered. \cite{kölle2024study}. 

In contrast to the aforementioned quantum variants of PPO, our proposed framework PPO-Q that illustrated in FIG. \ref{fig1} is capable of successfully handling diverse environments, encompassing both continuous and high-dimensional ones. More importantly, we have carried out successful training and inference on real superconducting quantum hardware based on PPO-Q.\par

\section{3. Hybrid Quantum-Classical networks of PPO-Q} 
\label{sec3}
\begin{figure*}[tbp]
    \centering
    \includegraphics[width=0.95\linewidth]{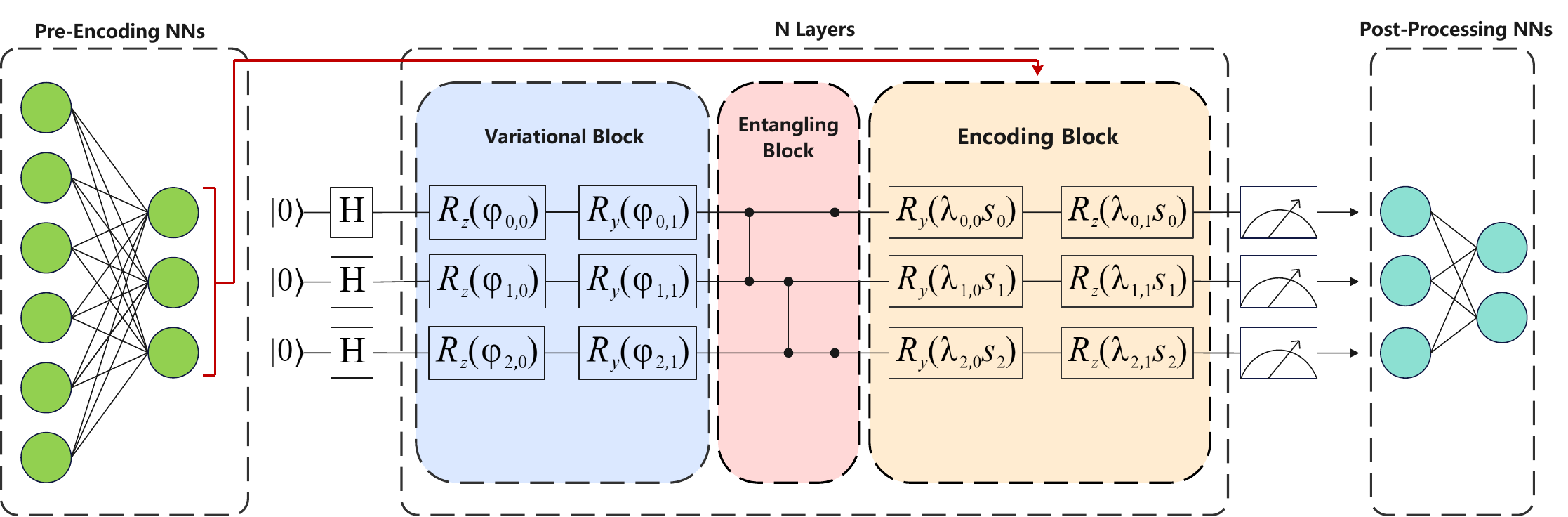}
    \caption{Hybrid quantum-classical networks of PPO-Q with three qubits as example.}
    \label{fig2}
\end{figure*}

To address complex control challenges, we devise a hybrid quantum-classical networks in the PPO-Q that encompasses pre-encoding NNs, PQC, and post-processing NNs, as illustrated in FIG. \ref{fig2}. The architecture of PQC adopts the widely-used hardware-efficient ansatz \cite{bib6}. It begins with H gates applied on each initial qubit, followed by $N$ quantum layers with each layer composed of a variational block, an entangling block, and an encoding block. Pauli-Z measurements are performed at the end of the quantum circuit. Single-qubit Y-rotation and Z-rotation gates are used for variational and encoding blocks, while two-qubit CZ gates are employed for entanglement generation. Since the two-qubit gates only act on adjacent qubits, they are particularly suitable for the physical qubits that form a closed one-dimensional chain. Within the overall networks, the pre-encoding NNs and the post-processing NNs are introduced to manage high-dimensional state encoding and action selection, respectively, thereby enhancing the functionality and performance of the PQC.
In the following, we provide a detailed description of how this hybrid networks works in practice.

\subsection{3.1. State Encoding}
\label{sec31}

For environments with a tractable-dimensional state space, the observed states $s$ are conventionally encoded directly into Y-rotation and Z-rotation gates with classical trainable scaling parameters $\lambda$. Owing to the periodicity of $2\pi$ inherent in rotation gates, rescaling function $\tanh(\lambda\cdot s)$ are employed to ensure that the scaled states lie within the range of $[-\pi, \pi]$ \cite{e25010093}. For environments with high-dimensional or continuous state space, nevertheless, the direct encoding of observed state $s$ is rather challenging due to the constraints of available quantum computing resources. In light of this, NNs are employed to conduct dimensionality reduction, thereby pre-encoding high-dimensional states. The output of the NNs function as state features that are utilized in the encoding block. The empirical results presented hereinafter demonstrate that the employment of pre-encoding NNs facilitates agents in handling complete tasks or attaining enhanced performance.

\subsection{3.2.  Measurement and Post-Processing}
\label{sec32}
The measurement strategies of PQCs play a pivotal role in determining the performance of the agents, as indicated in earlier research \cite{bib6, bib12, bib7}. In those approaches, the observables are meticulously customized and adjusted in accordance with each specific environment to ensure ultimate success. Moreover, the weighted observables with trainable wight $\omega_{a}$ to each action $a$ are employed to guarantee the variability of the output. Given that an $n$ qubits PQC with input state $s$, scaling parameters $\lambda$ in the encoding block, and rotation angles $\varphi$ in the variational block, the expectation value of the weighted observables can be expressed as
\begin{equation}
\left\langle O_a\right\rangle_{s, \boldsymbol{\theta}}=\left\langle 0^{\otimes n}\left|U(s, \boldsymbol{\varphi}, \lambda)^{\dagger} \hat{O}_a U(s, \boldsymbol{\varphi}, \lambda)\right| 0^{\otimes n}\right\rangle \cdot \omega_a,
\end{equation}
where $\hat{O}_{a}$ represent the observable operators, and $\boldsymbol{\theta}\equiv (\varphi, \lambda, \omega)$. In PPO-Q, we adopt the non-linear activation function softmax applied to the $\left\langle O_a\right\rangle_{s, \boldsymbol{\theta}}$, defining a SOFTMAX-PQC policy \cite{bib6} as
\begin{equation}
\pi_{\boldsymbol{\theta}}(a \mid s)=\frac{\mathrm{e}^{\beta\left\langle O_a\right\rangle_{s, \theta}}}{\sum_{a^{\prime}} \mathrm{e}^{\beta\left\langle O_{a^{\prime}}\right\rangle_{s, \theta}}},
\end{equation}
where $\beta$ is an inverse-temperature parameter.

Given that the identification of suitable observables is a time-consuming process, it would be highly challenging when attempting to accomplish tasks within novel and complex environments. To surmount this limitation, we introduce post-processing NNs to facilitate the handling of measurement outcomes. We apply Pauli-Z observable operators to all qubits to acquire comprehensive measurement data, which serves as the input for the post-processing NNs. In the case of discrete actions, the output of post-processing NNs is transformed into a probability distribution using the softmax function. The agent then selects its action according to this probability distribution. In the context of continuous actions, post-processing NNs are employed to calculate two variables of the beta distribution for each individual action \cite{pmlr-v70-chou17a}. This approach is in contrast to the widely adopted normal distribution, with the objective of attaining superior results. The action is selected by sampling from each beta distribution. In addition, concerning the critic values, the post-processing NNs output a single value that serves to represent the quality of a particular state for the agent.
\par
\subsection{3.3. Algorithm of PPO-Q}
\label{sec4}
PPO has consistently ranked among the most prominent online policy gradient methods, owing to its remarkable sample efficiency, simplicity, and advantages in training time \cite{bib44}. The core distinction of PPO-Q in comparison to PPO lies in the replacement of the NNs in the actor or critic component with our proposed hybrid quantum-classical networks. Given this fact, the algorithm of PPO-Q is nearly identical to that of PPO, with the sole difference residing in the policy network $\pi_{\hat{\theta}}$ or the value network $V_{\phi}$. During the training process using the PPO-Q, we have also adopted a lot of code-level optimization techniques \cite{bib51,bib52,bib53} based on the PPO, which have been thoroughly researched and proven effective. Algorithm \ref{algorithm1} shown below exhibits the comprehensive algorithmic representation in which hybrid quantum-classical networks serve as actors while the critics remain classical (Abbreviated as Hybrid Quantum Actor).  It bears resemblance to the scenarios where either critics are realized by utilizing hybrid quantum-classical networks meanwhile actors remaining classical (Hybrid Quantum Critic), or both actors and critics are constituted by hybrid quantum-classical networks (Full Quantum Actor-Critic).

\begin{algorithm}[H]
    \caption{PPO-Q}
    \label{algorithm1} 
    \begin{algorithmic}
        \State Initialize hybrid quantum-classical policy network $\pi_{\hat{\theta}}$ and classical value network $V_{\phi}$ with parameters $\hat{\theta}$ and $\phi$. 
        \State Initialize experience buffer.
        \State Set max interaction steps, batch size, number of epochs $K$, and mini-batch size. 
        \State Set hyperparameters $\gamma$, $\hat{\lambda}$, $\epsilon$, $c_1$, and $c_2$. 
        \State Determine the number of sampling iterations $M = \frac{\text{max interaction steps}}{\text{batch size}}$.
        
        \For{$m = 1$ to $M$}
        \State Collect batch size trajectories using policy $\pi_{\hat{\theta}}$:
        \For{each step $t$}
        \State Interact with environment, record $(s_t, a_t, r_t, s_{t+1}, \text{done})$
        \EndFor
        \State Compute discounted returns $R_t = r_t + \gamma r_{t+1} + \gamma^2 r_{t+2} + \dots$

        \State Compute temporal difference error $\delta_t = r_t + \gamma V_{\phi}(s_{t+1}) - V_{\phi}(s_t)$
        \State Compute Generalized Advantage Estimation (GAE) $\hat{A}_t = \delta_t + (\gamma \hat{\lambda}) \delta_{t+1} + (\gamma \hat{\lambda})^2 \delta_{t+2} + \dots$

        \State Store trajectories, returns and advantages in buffer
        
        \For{$k = 1$ to $K$}
        \State Shuffle buffer and split data into mini batches
        \For{each mini batch}
        \State Extract mini batch data: states, actions, returns, and advantages
        
        \State Normalize advantages at minibatch level  
        
        \State Compute policy ratio $r_t(\hat{\theta}) = \frac{\pi_{\hat{\theta}}(a_t | s_t)}{\pi_{\hat{\theta}_{\text{old}}}(a_t | s_t)}$

        \State Compute clipped objective $L^{\text{CLIP}}(\hat{\theta}) = \mathbb{E}_t \left[ \min \left( r_t(\hat{\theta}) \hat{A}_t, \text{clip}(r_t(\hat{\theta}), 1-\epsilon, 1+\epsilon) \hat{A}_t \right) \right]$

        \State Compute value function loss $L^{\text{VF}}(\phi) = \mathbb{E}_t \left[ (R_t - V_{\phi}(s_t))^2 \right]$
 
        \State Compute entropy bonus $H(\pi_{\hat{\theta}}) = \mathbb{E}_t \left[ -\sum_{a} \pi_{\hat{\theta}}(a|s) \log \pi_{\hat{\theta}}(a|s) \right]$
 
        \State Compute total loss $L = L^{\text{CLIP}}(\hat{\theta})- c_1 H(\pi_{\hat{\theta}})  + c_2 L^{\text{VF}}(\phi) $

        \State Update parameters $\hat{\theta}$ and $\phi$
        \EndFor
        \EndFor
        \EndFor
    \end{algorithmic}
\end{algorithm}

\section{4. Experiments}
\label{sec5}
We carry out a multitude of experiments based on the proposed PPO-Q within the standard benchmarking environments sourced from the OpenAI Gym library \cite{brockman2016openai}. Unless otherwise specified, the proposed hybrid quantum-classical networks are employed as the actors while keeping the original NNs as the critics in most experiments. Our experiments are arranged as follows. First, a comprehensive assessment is carried out to evaluate the effectiveness of the proposed hybrid quantum-classical networks, which include pre-encoding NNs, post-processing NNs, and the layer number of PQC. Subsequently, a performance comparison with classical NNs is presented across eight diverse environments, wherein four tasks feature continuous action spaces. Moreover, the evaluation of other essential schemes, i.e., the hybrid quantum critic and the full quantum actor-critic are also demonstrated. Finally, the proposed PPO-Q is implemented on real superconducting quantum devices supported by the Quafu quantum cloud service \cite{bib1}.

\subsection{4.1. Experiment Setup}
\label{sec51}
In the subsequent experiments, the simulation of PQCs is carried out with the assistance of TorchQuantum \cite{bib2}, whereas the classical NNs are implemented by means of PyTorch \cite{bib3}. The selected standard environments are illustrated in FIG. \ref{fig3} of Appendix A. Our experimental implementation encompasses four tasks featuring discrete action spaces, namely CartPole - v1, MountainCar - v0, Acrobot - v1, and LunarLander - v2. Additionally, we incorporate four tasks with continuous action spaces, including MountainCar(C) - v0, Pendulum - v1, LunarLander(C) - v2, and BipedalWalker - v3. The details regarding these environments are summarized in TABLE II, which presents the dimensions of the state space and the action space.


\begin{table}[htbp]
    \caption{The dimension of state space and action space in environments.}
    \footnotesize
    \begin{ruledtabular}
        \begin{tabular}{c|cccccccc}
            & CartPole & MountainCar & Acrobot & LunarLander & MountainCar(C) & Pendulum & LunarLander(C) & BipedalWalker \\
            \hline
            State Space & 4 & 2 & 6 & 8 & 2 & 3 & 8 & 24 \\
            Action Space & 2 & 3 & 3 & 4 & 1 & 1 & 2 & 4 \\
        \end{tabular}
        \label{tab1}
    \end{ruledtabular}
\end{table}

\subsection{4.2.  Evaluation of Hybrid Quantum-Classical Network}
\label{sec52}
To validate the effectiveness and generality of proposed hybrid quantum-classical network, we conduct a comprehensive examination of several critical components such as pre-encoding NNs and post-processing NNs on different representative environments. Furthermore, we investigate the influence of varying the layer number of PQC on QRL training. In addition, we elucidate the profound impact of diverse parameter initialization strategies on the hybrid network. These well-designed evaluations provide an in-depth understanding of the underlying mechanisms of proposed hybrid quantum-classical network.

\begin{figure*}[htbp]
    \centering
    \includegraphics[width=1\textwidth]{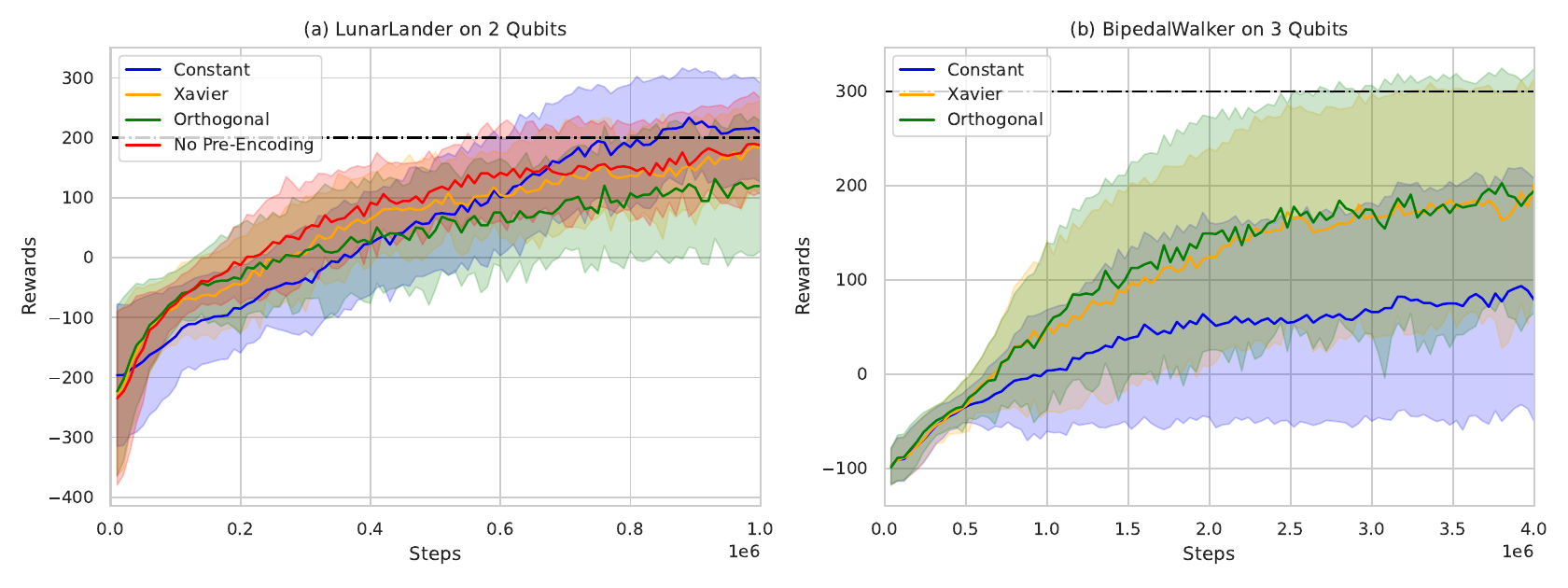}
    \caption{Learning curves of agents using pre-encoding NNs with different initialization strategies on LunarLander-v2 and BipedalWalker-v3. The solid line represents the mean of ten seeds, the shaded area indicates the standard deviation. The target reward of the environments are denoted by the dotted black line.}
    \label{fig4}
\end{figure*}
\textbf{Pre-encoding NNs}. In our framework, pre-encoding NNs are designed to handle various dimensions of state space, especially to address the challenge of high-dimensional state space that cannot be directly managed by prevalent data re-uploading schemes \cite{bib5}. As illustrated in FIG. \ref{fig4}, we conduct experiments on LunarLander-v2 and BipedalWalker-v3 environments, in which pre-encoding NNs play a crucial role in diminishing the dimensionality of the state space.
Specifically, for the LunarLander-v2 environment, the dimensionality is reduced from 8 to 2, while for the BipedalWalker-v3 environment, it is decreased from 24 to 3. The favorable results demonstrate the feasibility and effectiveness of dimensionality reduction via pre-encoding NNs.

Furthermore, the influence of different initialization strategies on the training effect is investigated. In the LunarLander environment, the agents with reduced dimensionality, which employ different initialization strategies, are all capable of successfully accomplishing the task. Although the agent without pre-encoding NNs also completed the task, the agent with Constant initialization achieves the highest rewards at the end of the entire training process. This outcome implies that quantum agents equipped with pre-encoding NNs possess the potential to exhibit superior performance compared to agents lacking pre-encoding processing.
Crucially, pre-encoding NNs endow quantum agents with the ability to effectively handle high-dimensional environments such as the BipedalWalker. As illustrated in FIG. \ref{fig4} (b), the agents initialized with the Xavier \cite{2010Understanding} and Orthogonal \cite{saxe2014exact} methods both demonstrate effective problem-solving capabilities within this environment, while the agent without pre-encoding NNs struggles to achieve this objective due to the constraints imposed by quantum hardware.

It is noteworthy that, within the BipedalWalker environment, the agent initialized with the Constant scheme fails to attain the target reward, which is in stark contrast to that observed in the LunarLander experiment. These empirical findings imply that a prudent choice of initialization methods may potentially augment performance to a certain degree.

\begin{figure*}[htbp]
\centering
\includegraphics[width=1\textwidth]{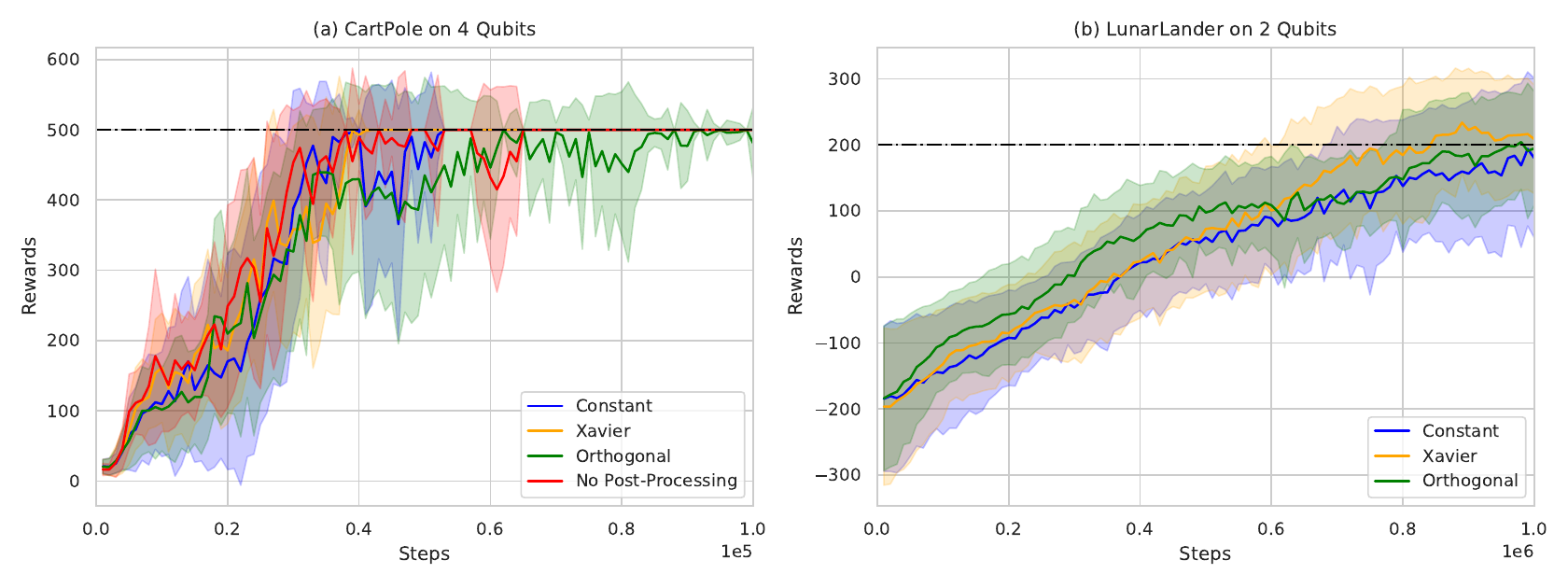}
\caption{Learning curves of agents using post-processing NNs with different initialization strategies on CartPole-v1 and LunarLander-v2. The interpretation of the lines is similar to that in FIG. \ref{fig4}.}
\label{fig5}
\end{figure*}
\textbf{Post-processing NNs.}  
In prior research concerning quantum policy methods \cite{bib6} and quantum actor-critic algorithms \cite{bib7}, the measurement of quantum circuits are customarily performed through meticulously designed observables. However, identifying suitable observables for each specific environment and deploying QRL algorithms in practical applications can be quite a challenge. Several studies \cite{bib8, bib9, bib10} have employed post-processing NNs layers; however, they have not managed to attain desirable performance. Here we conduct experiments on the CartPole-v1 and LunarLander-v2 environments to elucidate the effectiveness of post-processing NNs within proposed hybrid quantum-classical architecture.

We first show the implementation of the CartPole environment using a 4-qubit quantum circuit. CartPole is one of the most prevalently utilized standard environments, its observable has been previously designed and demonstrated to be effective. However, as shown in FIG. \ref{fig5} (a), the agent with post-processing NNs initialized using the Xavier method is the first to attain the target rewards, while the agent relying solely on observables ranks third. This indicates that the agent equipped with the post-processing step has the potential to achieve superior results. In the LunarLander environment, as shown in FIG. \ref{fig5} (b), we initially apply pre-encoding NNs to limit the circuit size to 2 qubits and subsequently adapt the measurement output through post-processing NNs. Each initialization strategy enables the post-processing NNs to fulfill the task; nevertheless, the agent initialized with the Xavier method achieves the top performance. Although some experiments regarding LunarLander have been conducted \cite{bib7, bib11}, which propose observables or post-processing steps for handling measurements, the post-processing NNs within our framework are more efficient and versatile. It is worth emphasizing again that a judiciously chosen initialization is crucial to achieve better performance.

\begin{figure*}[tbp]
\centering
\includegraphics[width=1\textwidth]{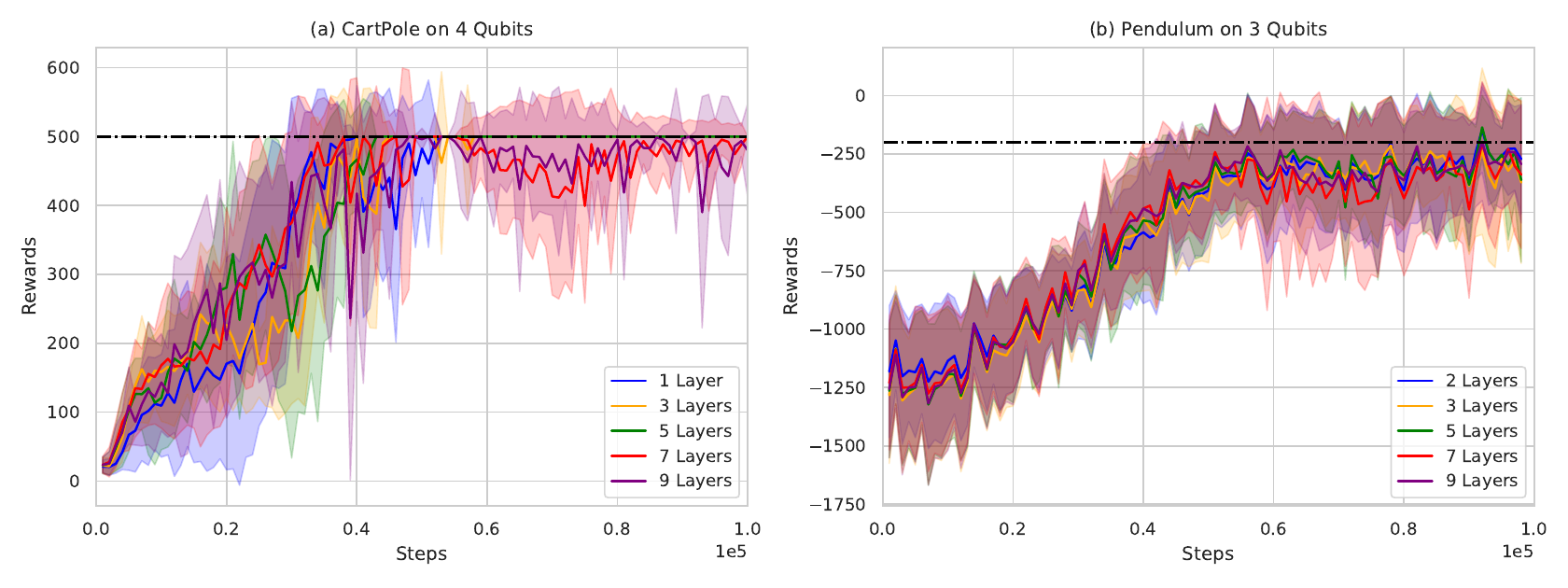}
\caption{Comparison of different number of variational quantum layers on CartPole-v1 and Pendulum-v1. The interpretation of the lines is similar to that in FIG. \ref{fig4}. We start recording the number of layers when the agent reaches the goal.}
\label{fig6}
\end{figure*}

\textbf{The layer number of PQC.}  
Increasing the layer number of PQC leads to an augmentation in the number of parameters, which usually enhances both the expressive power and performance. However, there exists a certain threshold, beyond which the performance will not be further improved \cite{bib12}. Here we present results on CartPole-v1 and Pendulum-v1 environments as shown in FIG. \ref{fig6}. It can be discerned, in the CartPole environment, that the agent attains optimal performance with five variational quantum layers, while additional layers lead to a decline in performance. Meanwhile, the agents with one layer and three layers achieve comparable results. In practice, we are inclined to select quantum circuits with a lower depth due to constrains on quantum computing resources.

The results of Pendulum, which is an environment with continuous action space, are depicted in FIG. \ref{fig6} (b). Unfortunately, the one-layer agent fails to accomplish the task. Although we are unable to discern a significant difference between the resulting curves, it is observed that the five-layer agent exhibits a marginally better performance than the seven-layer agent. Similar findings were also achieved in \cite{bib7} for the Pendulum with three-qubit circuits; however, they did not obtain consistently good performance across all different numbers of layers. This disparity may potentially arise from the differences in post-processing procedures, and our quantum architecture, in conjunction with its corresponding post-processing step, can lead to more dependable and stable performance.

\begin{table}[htbp]
    \caption{Setup of hybrid quantum-classical networks}
    \begin{ruledtabular}
        \begin{tabular}{lcccc}
            & CartPole & MountainCar & Acrobot & LunarLander \\
            \hline
            Pre-Encoding NNs & $\times$ & $\times$ & $\checkmark$ & $\checkmark$ \\
            Post-Processing NNs & $\checkmark$ & $\checkmark$ & $\checkmark$ & $\checkmark$ \\
            Initialization & ($\times$, Constant) & ($\times$, Constant) & (Constant, Xavier) & (Xavier, Constant) \\
            Qubits & 4 & 2 & 4 & 4 \\
            Number of Layers & 1 & 3 & 1 & 1 \\
            Quantum Parameters & 24 & 28 & 24 & 24 \\
            Total Actor Parameters & 32 & 34 & 60 & 72 \\
            \toprule
            & Pendulum & MountainCar(C) & BipedalWalker & LunarLander(C) \\
            \hline
            Pre-Encoding NNs & $\times$ & $\times$ & $\checkmark$ & $\checkmark$ \\
            Post-Processing NNs & $\checkmark$ & $\checkmark$ & $\checkmark$ & $\checkmark$ \\
            Initialization & ($\times$, Xavier) & ($\times$, Xavier) & (Orthogonal, Orthogonal) & (Constant, Orthogonal) \\
            Qubits & 3 & 2 & 4 & 4 \\
            Number of Layers & 2 & 3 & 1 & 1 \\
            Quantum Parameters & 30 & 28 & 24 & 24 \\
            Total Actor Parameters & 36 & 32 & 152 & 72 \\
        \end{tabular}
        \label{tab2}
    \end{ruledtabular}
\end{table}

\subsection{4.3. Results on Quantum Simulator}
We first conduct the implementation of PPO-Q on simulators and compare its performance with that of PPO across eight distinct environments. The setup of hybrid quantum-classical networks as policy networks is comprehensively detailed in TABLE \ref{tab2}. Meanwhile, the hyperparameters fine-tuned for the environments are presented in TABLE \ref{tab:hyperparameter} of Appendix A. The hyperparameters for PPO are either obtained from stable-baselines3-zoo \cite{bib13} or the hyperparameters for classical PPO are either obtained from stable-baselines3-zoo or are precisely fine-tuned by us in accordance with performance.
 Furthermore, we demonstrate that the proposed hybrid quantum-classical networks can seamlessly generalize to other quantum schemes, including the Hybrid Quantum Critic and the Full Quantum Actor-Critic. \par

\begin{figure*}[thbp]
\centering
\includegraphics[width=1\textwidth]{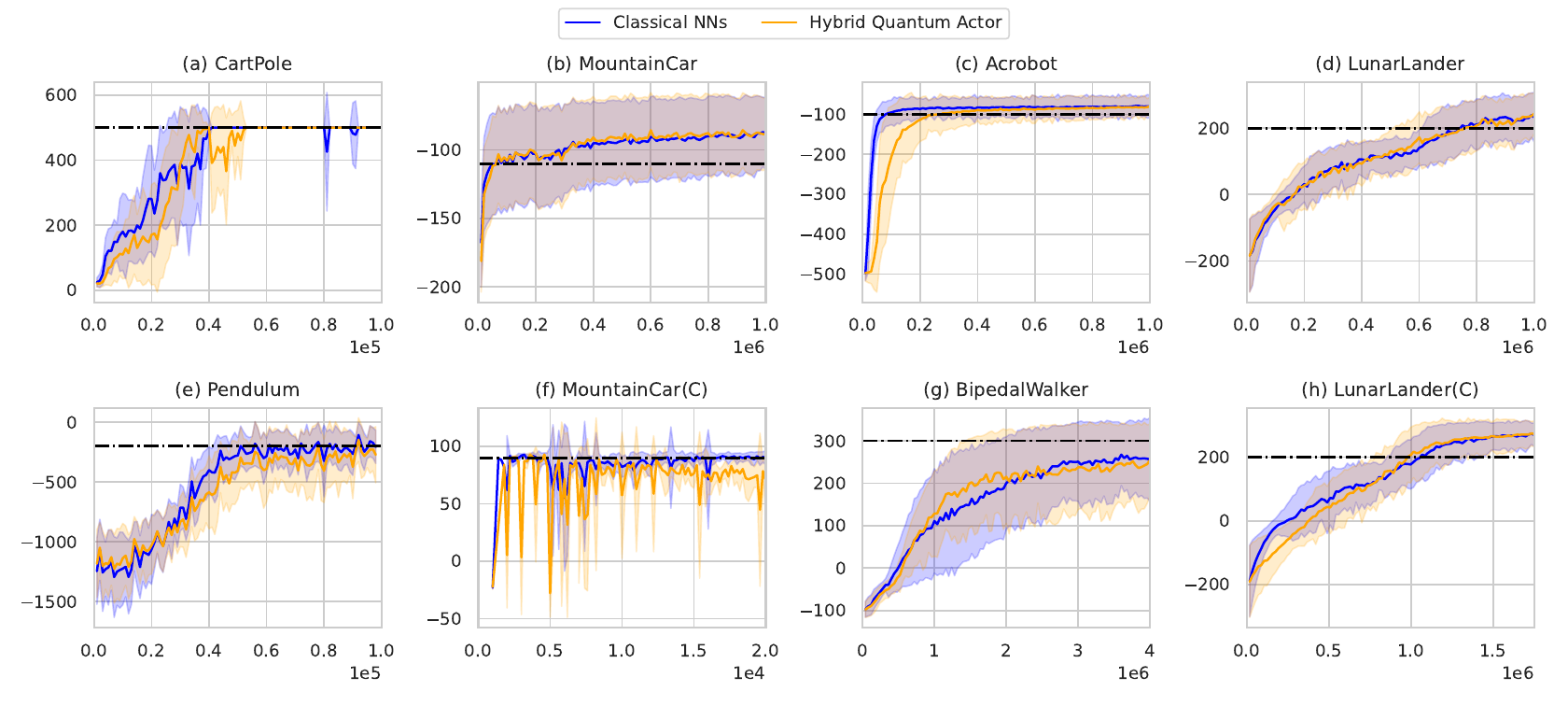}
\caption{Comparison between PPO-Q and PPO on eight environments. The orange line depicts the Hybrid Quantum Actor, while the blue line illustrates the agent with classical NNs. The solid line represents the mean of ten seeds, and the shaded area indicates the standard deviation. The target reward of the environments are denoted by the dotted black line.}
\label{fig7}
\end{figure*}

\textbf{Comparison with PPO.}
The comparison outcomes between PPO-Q and PPO are presented in FIG. \ref{fig7}, which can be observed that comparable performance is achieved across all eight environments. In four specific environments, i.e., MountainCar, BipedalWalker, LunarLander, and LunarLander(C), a more meticulous analysis discloses that not only are the mean performance tendencies similar to that of PPO but also the shaded standard deviation ranges are comparable. In MountainCar(C) and Pendulum environments, however, PPO-Q demonstrates certain subtly inferior performances. In MountainCar(C), the PPO-Q exhibits instability, whereas in Pendulum, it attains marginally lower rewards. 
In the remaining two environments, namely CartPole and Acrobot, the PPO-Q reaches the final milestone but at a relatively slower tempo in CartPole and exhibits a diminished speed during the initial reward acquisition phase in Acrobot. Despite the presence of minor discrepancies within certain environments, overall, it can be concluded that the PPO-Q achieves a performance level comparable to that of PPO. It validates that the proposed hybrid quantum-classical networks are bestowed with a stable and reliable capacity to execute tasks in a manner highly analogous to that of classical NNs. Most importantly, as distinctly presented in TABLE \ref{tab2}, PPO-Q has achieved such a performance with a substantially diminished number of training parameters.

\begin{figure*}[htbp]
\centering
\includegraphics[width=1\textwidth]{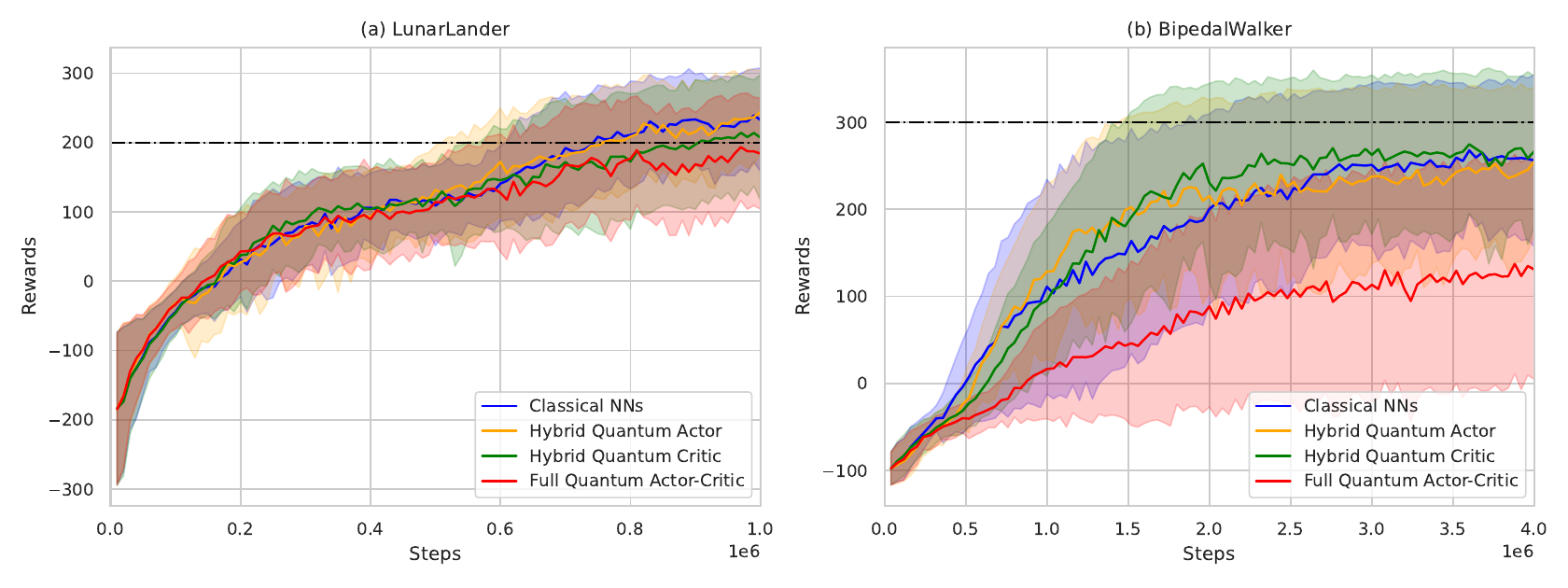}
\caption{Performance of all quantum schemes compared with classical NNs on LunarLander-v2 and BipedalWalker-v3. The green line represents the Hybrid Quantum Critic, while the red line indicates the Full Quantum Actor-Critic. Other lines are interpreted similarly to those in FIG. \ref{fig7}.}
\label{fig8}
\end{figure*}

\textbf{Other quantum schemes.}
In addition to the Hybrid Quantum Actor, we are also capable of readily implementing the Hybrid Quantum Critic and the Full Quantum Actor-Critic schemes within the framework of PPO-Q. The results of three quantum schemes in comparison with the classical one are presented for two disparate yet representative environments: LunarLander (characterized by a discrete action space) and BipedalWalker (characterized by a continuous action space). As depicted in FIG. \ref{fig8}(a), all quantum schemes can achieve the objective and exhibit performance comparable to that of classical NNs. Moreover, in FIG. \ref{fig8}(b), it can be observed that the Hybrid Quantum Critic persistently outperforms classical NNs in terms of rewards commencing from one million steps and throughout the remainder of the training process. Somewhat surprisingly, the Full Quantum Actor-Critic shows no superiority or bright spots. These findings validate the effectiveness of the Hybrid Quantum Actor or Critic schemes within the framework of PPO-Q. 

\subsection{4.4. Results on Real Quantum Devices}
We execute the PPO-Q algorithm on actual quantum devices supported by the Quafu quantum cloud service \cite{bib1}. Specifically, we selected a 105-qubit superconducting quantum chip named \textit{Dongling} for conducting the experiments. The basic information regarding \textit{Dongling} is provided in TABLE \ref{tab3}, and its topological structure is illustrated in FIG. \ref{fig11} of Appendix A.

\begin{table}[htbp]
    \caption{Basic information of superconducting quantum chip \it{Dongling}.}
    \footnotesize
    \begin{ruledtabular}
\begin{tabular}{ccccc}
        & Anharmonicity (MHz) & Qubit Frequency (GHz) & Readout Frequency (GHz)          & CZ Gate Fidelity\\ \hline
Average  & 0.210          & 4.305          & 6.800                       & 0.963                      \\
Min.     & 0.175         & 3.867           & 6.484                      & 0.880                       \\
Max.     & 0.223         & 4.809           & 7.136                      & 0.993                      \\ \toprule
        & T1 (us)           & T2 (us)           & Single Qubit Gate Duration (ns) & Single Qubit Gate Fidelity \\ \hline
Average & 59.536        & 37.887          & 31.39                      & 0.999                      \\
Min.     & 14.080            & 10.440           & 30.00                         & 0.993                      \\
Max.     & 96.500          & 124.930          & 45.00                         & 0.9999                  
\end{tabular}
        \label{tab3}
    \end{ruledtabular}
\end{table}

\begin{figure*}[tbp]
\centering
\includegraphics[width=1.0\textwidth]{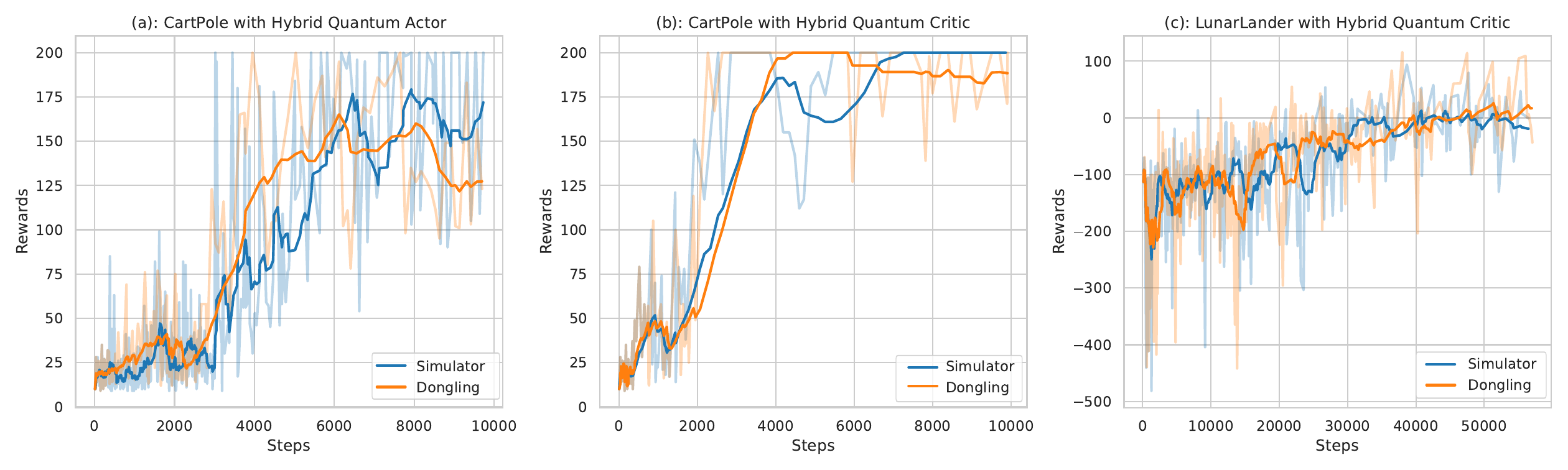}
\caption{Training performance of CartPole-v0 and LunarLander-v2 with comparison on both simulator and \textit{Dongling}. The solid line represents the moving window average of the original training process.}
\label{fig9}
\end{figure*}

We initially choose CartPole as the representative discrete-action environment and train both the Hybrid Quantum Actor and Critic schemes on the \textit{Dongling} device. As illustrated in FIG. \ref{fig9} (a) and (b), both complete the entire training process and attain a performance level comparable to that of the simulator, with the Hybrid Quantum Critic scheme being particularly notable. In light of the results obtained from CartPole and taking into account the time limitations of the available quantum hardware resources, we exclusively adopt the Hybrid Quantum Critic scheme for training the representative continuous-action environment, LunarLander, and the results are illustrated in FIG. 8(c). Our results imply that PPO-Q is capable of accomplishing satisfactory training on the current noisy quantum hardware, and can demonstrate a performance that is comparable to that of the simulator. \par

\begin{figure*}[t]
\centering
\includegraphics[width=1.0\textwidth]{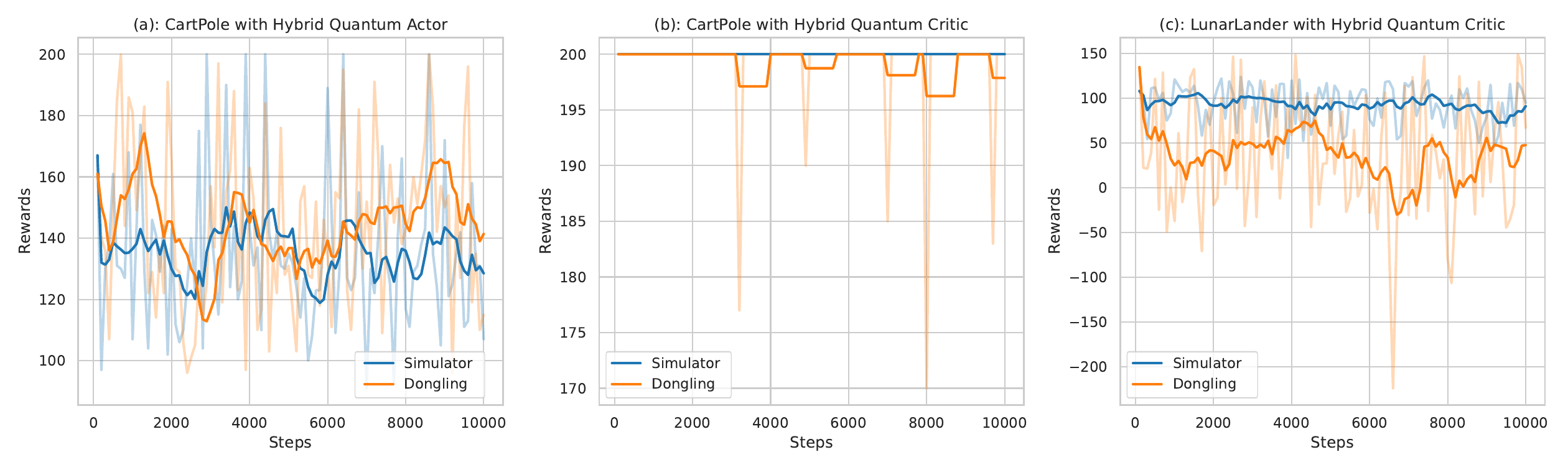}
\caption{Inference of the trained agents on both simulator and \textit{Dongling}. The solid line represents the moving window average of the original inferring process.}
\label{fig10}
\end{figure*}

The inference processes of the agents trained on the \textit{Dongling} are depicted in FIG. \ref{fig10}. It is observable that all agents are capable of inferring satisfactory results. In the CartPole with Hybrid Quantum Actor scheme, as shown in FIG. \ref{fig10} (a), the actor can even obtain better rewards on the \textit{Dongling} than in the simulator. Meanwhile, as depicted in FIG. \ref{fig10} (b) and (c), the actor within the Hybrid Quantum Critic scheme exhibits a relatively inferior performance on the \textit{Dongling} compared to that of the simulator. Such a result is relatively understandable as, in the Hybrid Quantum Actor scheme, we directly train the policy network on the real quantum hardware, which renders the actor more robust to the noise of the quantum hardware.
\par

\section{5. Conclusion}
\label{sec6}
In this work, we put forward PPO-Q, a hybrid quantum-classical reinforcement learning framework, intending to address complex control problems. The core of PPO-Q is a hybrid quantum-classical network that consists of three principal components: pre-encoding NNs, parameterized quantum circuits, and post-processing NNs. The pre-encoding NNs proficiently handle high-dimensional states, facilitating the successful training process and rendering the approach applicable in real-world scenarios. The PQCs are engineered to be hardware-efficient, guaranteeing that they can yield satisfactory results even when implemented on real quantum devices. The post-processing NNs obviate the necessity for the time-consuming selection of quantum observables and can transform measurement outcomes into either discrete or continuous actions, or specific critic values, thereby enhancing the architecture's flexibility.

Experimental results demonstrate that PPO-Q can effectively solve eight distinct environments and exhibit performance comparable to that of PPO while possessing notably fewer training parameters. Notably, it achieves a substantial advancement by successfully resolving the BipedalWalker environment, which had hitherto remained unsolved in the realm of QRL. Remarkably, the Hybrid Quantum Critic scheme within the BipedalWalker environment exhibits better performance compared to that of classical NNs. Moreover, we deploy PPO-Q on real quantum devices via the Quafu quantum cloud platform. The agents managed to successfully tackle the CartPole-v0 and LunarLander-v2 environments on a 105-qubit device \textit{Dongling}. The experimental findings imply that PPO-Q is highly effective in resolving complex problems, including those involving high-dimensional states and continuous actions.

Looking forward, PPO-Q can be further expanded to address even more intricate and demanding problems, regardless of whether they stem from OpenAI Gym environments or real-world applications. With the rapid advancement of current quantum chip technology \cite{google, ustc, IBM}, we are convinced that PPO-Q holds the potential to assist us in exploring the application potential of quantum computing in the field of artificial intelligence.

\begin{acknowledgments}
{\bf Acknowledgments:} 
The authors acknowledge Haifeng Yu, Yirong Jin, Heng Fan, and Yulong Feng for valuable discussions. We especially thank the Quafu Quantum Computing Hardware Team for technical support.
This work is supported by the Natural Science Foundation of China (Grant No. 92365206, No. 92476205, No. 92365111), the Beijing Natural Science Foundation (Grant No.Z220002), and the Innovation Program for Quantum Science and Technology (Grant No. 2021ZD0302400). 

\end{acknowledgments}

{\bf Code availability:} The code related to this work is available in \cite{PPOQ}.

\bibliographystyle{apsrev4-1}
\bibliography{refs}

\newpage   
\clearpage 
\appendix 
\onecolumngrid

\section{Appendix A}

\begin{table}[htbp]
    \caption{\textbf{\small{Hyperparameters of different environments}}}
    \begin{ruledtabular}
        \begin{tabular}{lcccc}
            & CartPole & MountainCar & Acrobot & LunarLander \\
            \hline
            Num. Epochs & 20 & 4 & 4 & 4 \\
            Minibatch Size & 256 & 64 & 64 & 64 \\
            Number of Actors & 8 & 16 & 16 & 16 \\
            Buffer Size & 256 & 256 & 4096 & 16348 \\
            Learning Rate of Actor & 1e-2 & 3e-3 & 3e-3 & 3e-3 \\
            Learning Rate of Critic & 1e-3 & 3e-4 & 3e-4 & 3e-4 \\
            Critic Hidden Neurons & (64, 64) & (64, 64) & (64, 64) & (64, 64) \\
            
            Discount ($\gamma$) & 0.98 & 0.99 & 0.99 & 0.999 \\
            GAE Parameter ($\hat{\lambda}$) & 0.8 & 0.98 & 0.98 & 0.98 \\
            \toprule
            & Pendulum & MountainCar(C) & BipedalWalker & LunarLander(C) \\
            \hline
            Num. Epochs & 10 & 10 & 7 & 4 \\
            Minibatch Size & 64 & 256 & 64 & 64 \\
            Number of Actors & 4 & 1 & 16 & 16 \\
            Buffer Size & 4096 & 8 & 16348 & 16348 \\
            Learning Rate of Actor & 1e-2 & 7.77e-4 & 4e-3 & 3e-3 \\
            Learning Rate of Critic & 1e-3 & 7.77e-5 & 1e-3 & 3e-4 \\
            Critic Hidden Neurons & (64, 64) & (64, 64) & (64, 64) & (64, 64) \\
            
            Discount ($\gamma$) & 0.9 & 0.999 & 0.99 & 0.999 \\
            GAE Parameter ($\hat{\lambda}$) & 0.95 & 0.9 & 0.95 & 0.98 \\
        \end{tabular}
        \label{tab:hyperparameter} 
    \end{ruledtabular}
\end{table}
\bigskip

\begin{figure}[htbp]
    \centering
    \includegraphics[width=0.9\linewidth]{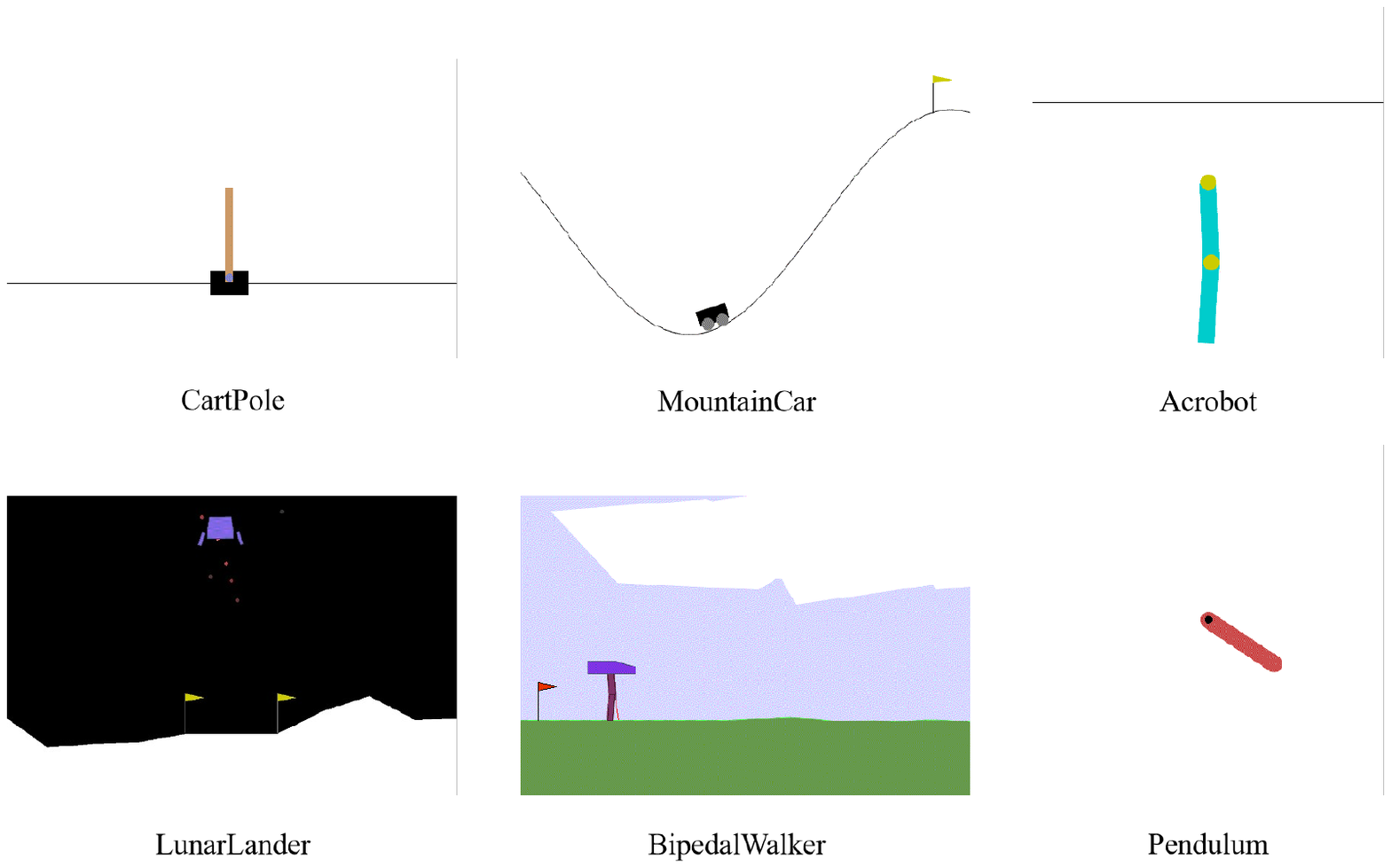}
    \caption{Benchmark environments from OpenAI Gym \cite{brockman2016openai}.}
    \label{fig3}
\end{figure}

\begin{figure}[!ht]
    \centering
    \includegraphics[width=0.71\linewidth]{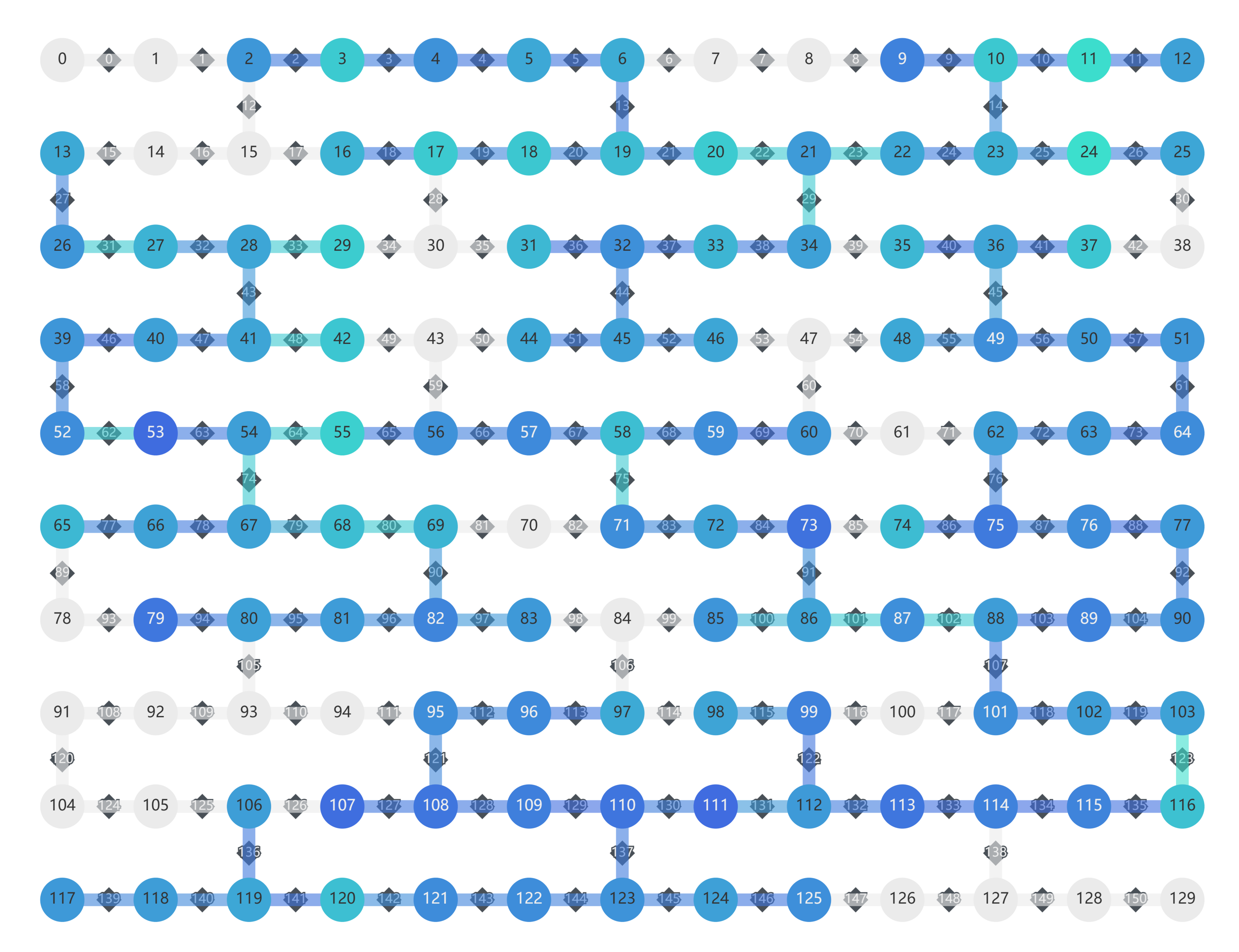}
    \caption{Topological structure of superconducting quantum chip \it{Dongling}.}
    \label{fig11}
\end{figure}











\end{document}